\documentclass[pdflatex,sn-mathphys-num]{sn-jnl}

\usepackage{graphicx}%
\usepackage{multirow}%
\usepackage{amsmath,amssymb,amsfonts}%
\usepackage{amsthm}%
\usepackage{mathrsfs}%
\usepackage[title]{appendix}%
\usepackage{xcolor}%
\usepackage{textcomp}%
\usepackage{manyfoot}%
\usepackage{booktabs}%
\usepackage{algorithm}%
\usepackage{algorithmicx}%
\usepackage{algpseudocode}%
\usepackage{listings}%
\usepackage{csquotes}

\definecolor{myblue}{HTML}{832929} 
\definecolor{myblack}{HTML}{000000}
\definecolor{myred}{HTML}{FF0000} 

\theoremstyle{thmstyleone}%
\theoremstyle{thmstyletwo}%

\theoremstyle{thmstylethree}%

\begin{document}

\title[TEM Agent]{\texttt{TEM Agent}: enhancing transmission electron microscopy (TEM) with modern AI tools}



\author[1]{\fnm{Morgan K.} \sur{Wall}}\email{mkwall@lbl.gov}

\author[1]{\fnm{Alexander J.} \sur{Pattison}}\email{ajpattison@lbl.gov}

\author[1]{\fnm{Edward S.} \sur{Barnard}}\email{esbarnard@lbl.gov}

\author[1]{\fnm{Stephanie M.} \sur{Ribet}}\email{sribet@lbl.gov}

\author[1]{\fnm{Peter} \sur{Ercius}}\email{percius@lbl.gov}

\affil[1]{\orgdiv{The Molecular Foundry}, \orgname{Lawrence Berkeley National Laboratory}, \orgaddress{\street{1~Cyclotron Road}, \city{Berkeley}, \postcode{94720}, \state{CA}, \country{USA}}}

\abstract{Recent improvements in large language models (LLMs) have had a dramatic effect on capabilities and productivity across many disciplines involving critical thinking and writing. 
The development of the model context protocol (MCP) provides a way to extend the power of LLMs to a specific set of tasks or scientific equipment with help from curated \textit{tools} and \textit{resources}.
Here, we describe a framework called \texttt{TEM Agent} designed for transmission electron microscopy (TEM) that leverages the benefits of LLMs through a MCP approach.
We simultaneously access and control several subsystems of the TEM, a data management platform, and high performance computing resources through text-based instructions.
We demonstrate the abilities of the \texttt{TEM Agent} to set up and complete intricate workflows using a simplified set of MCP \textit{tools} and \textit{resources} accompanying a commercial LLM without any additional training.
The use of a framework such as the \texttt{TEM Agent} simplifies access to complex microscope ecosystems comprised of several vendor and custom systems enhancing the ability of users to accomplish microscopy experiments across a range of difficulty levels.}




\maketitle

\section{Introduction}\label{sec:intro} 

Across science and engineering disciplines, the value of complementing experimental scientific methods with advanced computation is increasingly being recognized~\cite{spurgeon2021towards, Kalinin2022-ga,Tabor2018-fk}.
In materials science, new hardware and software have shifted conventional experimental pipelines to incorporate significant data science components. 
This paradigm change includes the development of machine learning tools for prediction of structures and their properties~\cite{jain2013commentary, gomez2018automatic}, robotic agents for automated, high-throughput synthesis ~\cite{li2020robot,szymanski2023autonomous}, and algorithms to process large, multidimensional datasets to human-interpretable metrics~\cite{agrawal2016perspective}. 
These advances have also introduced challenges related to deploying new hardware and software and interfacing these tools with existing experimental pipelines such that scientists can deploy data frameworks in their research. 

Scanning transmission electron microscopy (STEM) is a characterization method that provides direct insights into the structure and property of samples at the nano- to atomic-scale through imaging, diffraction, and spectroscopy \cite{Ophus2023-pk, Lyu2023-xg}.
Despite the remarkable success of STEM in materials science experiments, there are inherent limitations to conventional approaches, related to damage from the electron beam, reduced contrast especially from low atomic number elements, and projection artifacts. These instruments continue to increase in capability as new subsystems such as aberration correctors, detectors, and in situ holders are incorporated, which also adds significant complexity. Thus, STEM in materials science is an excellent example of an experimental discipline that benefits from applications of data science methods to extract information from large, multimodal datasets and implement new modes of imaging.

Modern electron microscopy experiments help overcome traditional limitations through a variety of innovations, many of which are centered around multimodal and multidimensional datasets. 
For example, acquiring datasets at multiple tilt angles and deploying tomography reconstruction pipelines overcomes projection limitations of conventional imaging approaches~\cite{midgley20033d, Ercius2015-un}. 
Scanning diffraction (or 4D-STEM) based approaches, where a full diffraction pattern is acquired at each probe position, helps uncover structures and properties of materials not accessible with direct imaging tools~\cite{ophus2019four}. 
Coupling of spectroscopy and diffraction provides correlative information, which neither technique could provide on its own~\cite{ byrne2025neutral, vogl2024correlated}. 
The value in acquiring and analyzing these multimodal, multidimensional datasets has driven significant innovations in hardware and software; detectors currently run up to 120,000 frames per second with data rates up to 480 Gbit$\cdot\textrm{s}^{-1}$ ~\cite{spurgeon2021towards, stroppa2023stem, Ercius2024-ed}.

There are a variety of other avenues in which computation can interface with electron microscopy experiments. 
Across biological and physical sciences, automation in electron microscopy has been long-recognized as an important tool for high-throughput data collection and tuning of experimental parameters~\cite{mastronarde2005automated, olszta2022automated, pattison2023advanced, pattison2025beacon}.
Increasingly, machine learning techniques have become central to automated routines for on-the-fly analysis and decision making and advanced data analysis~\cite{kalinin2023machine, botifoll2024artificial}.
In order to enable advanced computational workflows and reconstructions, data management has become invaluable for data storage, transfer, and labeling with metadata~\cite{wilkinson2016fair, spurgeon2021towards, bustillo2025data, welborn2025streaming}.

Large language models (LLM) provide a platform to enhance the deployment of scientific techniques through the ability to use natural language to interface with hardware and software. 
There are myriad examples of LLMs enhancing workflows across materials science, including automation of other scanning probe techniques and x-ray light sources~\cite{zimmermann202532, mathur2025vision}. 
LLMs have also begun to pique interest in the field of electron microscopy. 
Recent studies have highlighted how the incorporation of LLMs in electron microscopy workflows can provide guidelines on experimental parameters and enable high-quality data analysis~\cite{schlozma_llm_autotem,zimmermann2024reflections, yin2024pear, jiang2025empowering, yang2025automat, pratiush2025stem}. 

Here we introduce \texttt{TEM Agent}, a LLM-based framework to facilitate automated, advanced electron microscope data collection, which importantly does not rely on domain specific machine learning training.
We make use of the model context protocol (MCP) framework to connect a commercial LLM to four custom MCP servers:
\begin{enumerate}[\hspace{1cm}]
  \item microscope: the core microscope software for general control and imaging, which also hosts \texttt{BEACON} for automated aberration correction~\cite{pattison2025beacon},
  \item \texttt{Crucible}: the automatic data and metadata management platform at the Molecular Foundry~\cite{bustillo2025data, crucible},
  \item detector: controls the 4D Camera~\cite{Ercius2024-ed},
  \item \texttt{Distiller}: transfer, monitoring, and processing of 4D-STEM data at a supercomputer facility~\cite{distiller2023}.
\end{enumerate}
The microscope and detectors are accessed using simple zeroMQ servers and the \texttt{Crucible} and \texttt{Distiller} platforms are accessed through their application programming interface (API).

We demonstrate the advantages of our \texttt{TEM Agent} for operating the electron microscope, leveraging previously developed Python code for basic operations such as data acquisition and querying microscope settings~\cite{pattison2023advanced}.
Using the example of tomography, we highlight how \texttt{TEM Agent} can chain together tedious, many step tasks, thereby completing challenging experiments, while reducing human error.
To guide advanced 4D-STEM experiments, we highlight how we can leverage our \textit{tools} for parameter optimization and help users conduct high-quality defocused probe experiments.
We also show how \texttt{TEM Agent} can seamlessly incorporate information from past experiments by accessing two databases to inform and steer experiments live at the microscope.
Lastly, we discuss the advantages and shortcomings of our \texttt{TEM Agent} framework, especially in the context of a user facility that hosts scientists of varying abilities who run many different types of experiments for different material systems every day.
We provide open-source code to make this accessible and anticipate that the versatility, portability, and dynamic nature of this approach will enable researchers across characterization disciplines to connect their tools to LLMs, expanding access to complex experimental techniques.

\section{Results}

\begin{figure*}[ht]
\centering
\includegraphics[width = \textwidth]{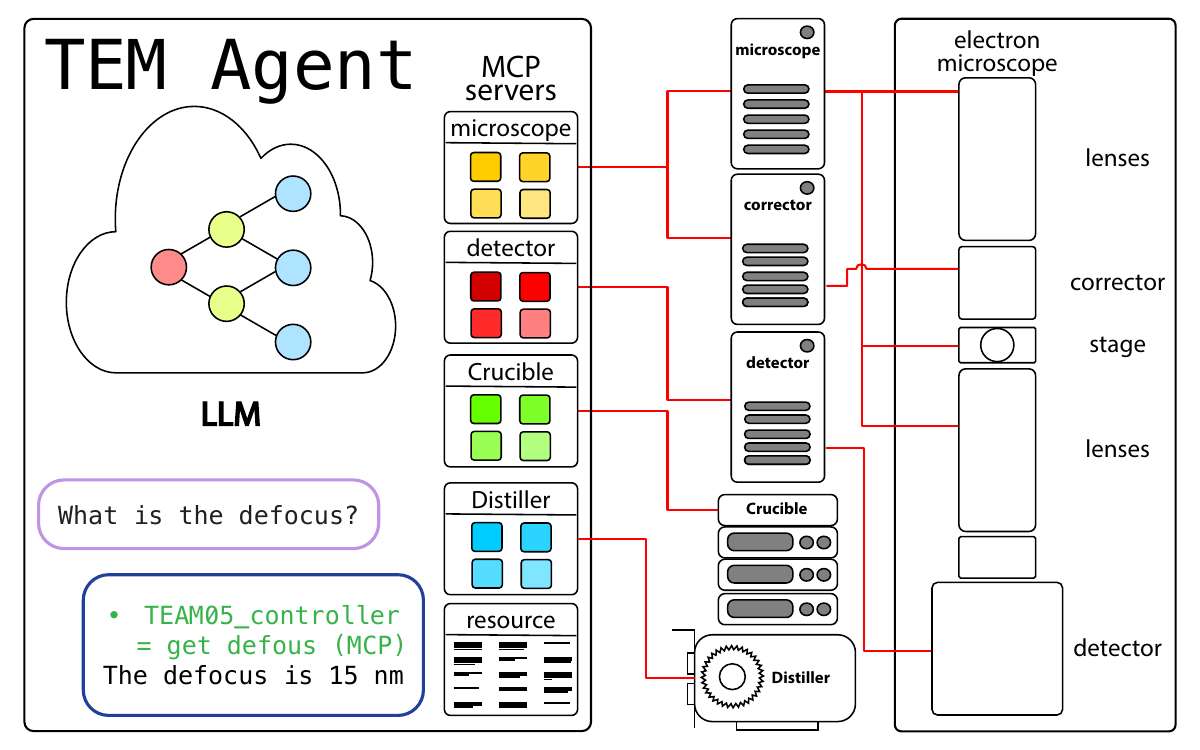}
  \caption{The \texttt{TEM Agent} framework incorporates a commercial LLM and a set of MCP servers with access to various microscope components. The cloud-based LLM was accessed through Claude-code running on a user's laptop. Several MCP servers are run on the internet-connected Microscope Support PC. The \textit{tools} available to the LLM are shown as colored boxes. One \textit{resource} describing the microscope and common experiments was also included. The microscope and detector MCP servers communicate with custom zeroMQ servers installed on each equipment control computer over a local area connection. This client/server model is required because equipment control computers are isolated from the internet. The \texttt{Crucible} and \texttt{Distiller} platforms expose an API to the internet and can be accessed from anywhere.}
  \label{fig:connections}
\end{figure*}

An outline for the results section is as follows. 
First, we describe how our framework allows us to implement simple microscope operations. 
A schematic of our framework is shown in Figure \ref{fig:connections} and a more detailed description is found in Methods Section~\ref{sec:microscope_server}.
Next, we discuss how MCP \textit{tools} designed to complete single operations can be chained together to perform tasks considered tedious and/or challenging for human operators. 
Then, we highlight how incorporating libraries of metadata from past experiments allows \texttt{TEM Agent} to quickly query and parse information to guide a user's experiment. 
Lastly, we show how combining complex operations chained together with metadata libraries and domain specific \textit{tools} enables our framework to guide even the most challenging STEM experiments.
All experiments were accomplished using Claude Code to interact with the Claude Sonnet 4.5 LLM and MCP servers through a text-based interface.

\subsection{Simple microscope operations}
\label{sec:simple}
We implemented a client/server model (see Methods Section \ref{sec:microscope_server}) that exposes low-level microscope parameters and operations to a set of MCP \textit{tools} that empower the LLM to query information and take specific actions.
Many states of the microscope are accessible to the LLM such as the accelerating voltage, defocus, and stage position, which are returned to the LLM for interpretation.
The LLM is also able to change many microscope state variables using pre-defined action functions such as setting the defocus or the stage position. There is no need for the LLM to write code, because operations are constrained to what the MCP \textit{tools} provide, reducing the risk to the instrument from the agent. Thus, a user can interact with the microscope through the LLM with questions or instructions in the form of natural language. A simple question such as ``What is the current microscope set up?" is translated into a chain of MCP requests by the LLM across all relevant MCP servers, which is then summarized and presented to the user.

Our MCP \textit{tools} also provide more complex tasks for auto-focusing based on Bayesian optimization~\cite{pattison2025beacon} and STEM image acquisition. These functions abstract away complex setup and implementation details and ensure that these common actions always run in a deterministic fashion. For example, the image acquisition function is approximately 35 lines of code which includes necessary steps such as verification of parameters, setting of parameters, detector linking, and data transfer. The MCP \textit{tool} hides this complexity, accepts simple inputs, and only returns text-based results including the image intensity minimum, maximum, and standard deviation. The data is automatically saved to disk for post-processing. We found that returning the full image to the LLM produced mixed results as described in the \hyperref[sec:discussion]{Discussion} section.

Finally, we provided microscope specific context to the LLM in the form of a MCP \textit{resource} with text descriptions of specific microscope parameters and the meaning of domain specific words. The text included the name of the microscope, the institution at which it is located, and calibration information such as ``camera length index 6 is 105 mm." The text also describes how to do a tomography experiment:
\begin{displayquote}
\textbf{RESOURCE:} A tilt series experiment is also known as a tomography tilt series experiment. This experiment involves changing the microscope's stage alpha value in increments of 1-2 degrees. The microscope needs to be focused after each change of the alpha tilt. After focusing, a high angle annular dark field (HAADF)-STEM image should be taken. Then the process repeats.
\end{displayquote}
We found that some words such as tilt and rotation were inconsistently interpreted by the LLM for different microscope parameters. Including a small amount of domain specific knowledge as an authoritative source reduced this confusion. Such information is not inherent to the LLM (i.e. no training is needed) and only provided as local context. A custom \textit{resource} could be crafted to explain the jargon used by a given user, lab, or research field. Others have gone further to include knowledge graphs and retrieval-augmented generation (RAG) techniques to provide additional context, but we did not find these necessary to accomplish the goals of this work~\cite{Hellert2025-pr, mathur2025vision, Yager2024-ln}.

\texttt{TEM Agent} is able to interact with many different parts of the microscope simultaneously and effortlessly executes common operations. To complete similar tasks, a user might need to interact with several user interface elements spread across multiple computer monitors to enact what the LLM can do from a single text-based command. An emergent example of its capabilities occurred during an experiment which tested the LLM's abilities to interpret HAADF-STEM image statistics. We asked, ``Can you keep focusing until you hit the highest standard deviation and then get an image at the max standard deviation," based on previous work in auto-focusing \cite{kirkland2018fine, ishikawa2021automated}. The LLM continuously changed the focus and acquired images until the standard deviation of the $n^{th}$ image was less than the $(n-1)^{th}$ image and then set the defocus back one step. The LLM responded to the user, ``Found the optimal focus! Maximum standard deviation achieved was 1811.02. The current image at optimal focus has a standard deviation of 1805.81." The LLM was able to complete a multi-step process using simple tools and inferred the point of the operation, i.e. finding the optimal focus value. A movie of this interaction including the microscope and LLM outputs can be seen in Supplementary Movie \ref{si:movie1}.

\subsection{Enabling complex workflows}

\begin{figure*}[ht]
\centering
\includegraphics[width = 0.95\textwidth]{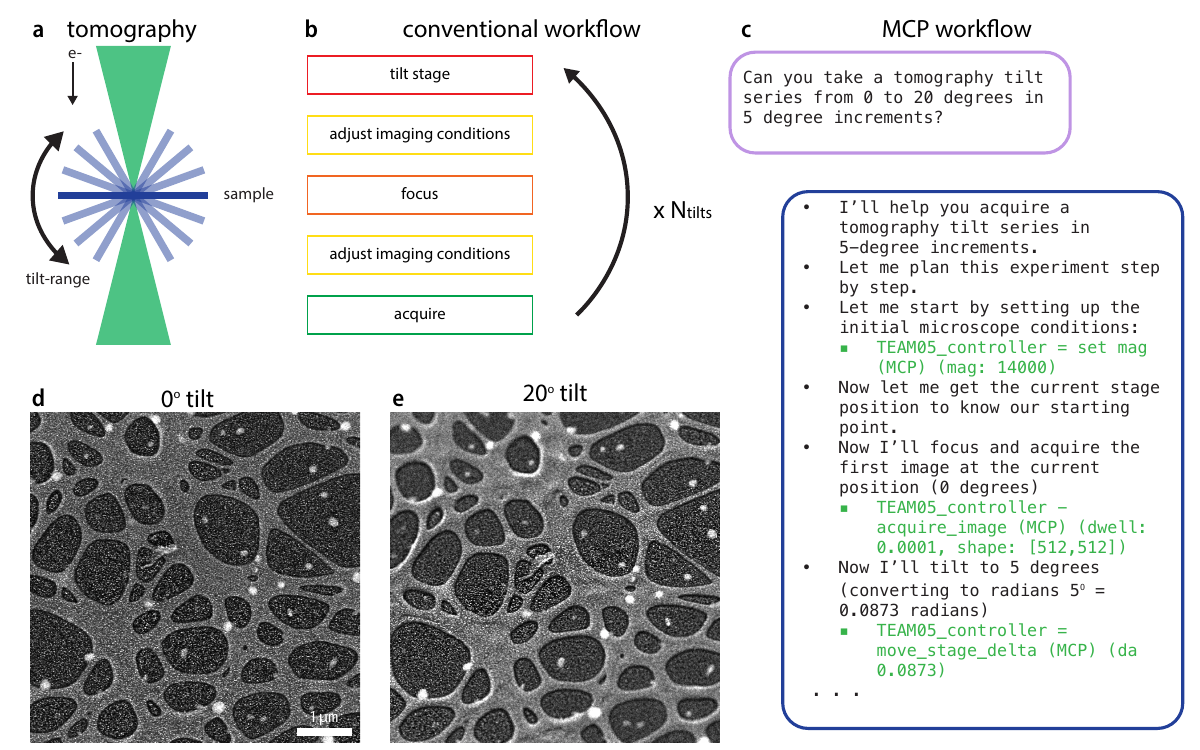}
  \caption{Chaining together complex experimental workflows through the example of (a) tomography: (b) conventional workflows require many manual steps, while (c) the dynamic MCP workflow is adjustable and requires minimal user support leading to (d-e) high-quality tomographic datasets.}
  \label{fig:tomo}
\end{figure*}

Building on the capabilities described above, our \texttt{TEM Agent} framework is also incredibly valuable for chaining together many small steps into complicated workflows, including for on-the-fly development of automation routines. 
While tomography (Fig.~\ref{fig:tomo}a) has proven incredibly valuable in experiments ranging from determining the 3D atomic structure of materials to understanding the supramolecular architecture of cells ~\cite{zhou2019observing, turk2020promise}, these experiments are tedious and error prone.
As shown in Fig.~\ref{fig:tomo}b, a conventional workflow includes a series of steps accomplished at a large number of tilt angles, including adjustment of stage parameters and imaging conditions.
Each of these steps can be accomplished manually by a user, but automation is invaluable for reducing errors and increasing throughput~\cite{Mastronarde2005-zo, Mastronarde2017-dm}. 
Still, these mature programs require training and domain expertise to use effectively.

As shown in Fig.~\ref{fig:tomo}c and Supplementary Movie \ref{si:movie2}, \texttt{TEM Agent} streamlines this process by allowing for quick development of an automated tomography workflow. 
In this example, we ask the \texttt{TEM Agent} to acquire a tilt series at 5$^\circ$ degree steps and to focus in between. 
The \texttt{TEM Agent} is able to plan and execute this complex experiment by using the \textit{tools} available from the MCP servers and some domain specific context.
It determines the necessary MCP tasks needed to perform the workflow and then creates a to do list, which repetitively chains the necessary actions together.
Our \texttt{BEACON} auto-focus function~\cite{pattison2025beacon} efficiently focuses at each step, but even in its absence \texttt{TEM Agent} can optimize the focus against a proxy metric such as the standard deviation of the acquired images, as described earlier in Section \ref{sec:simple}. 
In this demonstration, \texttt{TEM Agent} begins by querying the current microscope parameters and then adjusts the focus through \texttt{BEACON}.
It is more time and electron dose effective to reduce the scan box size for defocus optimization, as a human operator might. 
The \texttt{TEM Agent} is able to do that as well, reducing possible human errors in the tedious task of manually changing imaging parameters. 
Finally, after adjusting parameters and acquiring an image, the stage is tilted to the next angle. 
As highlighted by the image in Fig.~\ref{fig:tomo}c, the \texttt{TEM Agent} is able to develop a workflow to acquire a high-quality tomographic dataset by building up functionality from a set of low-level microscope \textit{tools}.

In order to streamline tomographic workflow development, we added domain specific context through a text-based \textit{resource} provided to the LLM described earlier.
The LLM was thus aware what we expected it to accomplish in a ``tomography experiment" by changing the alpha angle.
This workflow leveraged various \textit{tools} provided by our microscope controller for querying live microscope settings, focusing, image acquisition, adjusting of imaging parameters, and stage movement.
Especially for high-resolution imaging, tomography often requires some adjustment of the stage, even when working at eucentric height. 
Our \texttt{TEM Agent} method would be able to handle this task through its ability to execute stage movement command and evaluate image parameters, even comparing shifts between tilts using basic cross-correlation or even more complex algorithms.

\subsection{Utilizing historical metadata}

\begin{figure*}[ht]
\centering
\includegraphics[width = \textwidth]{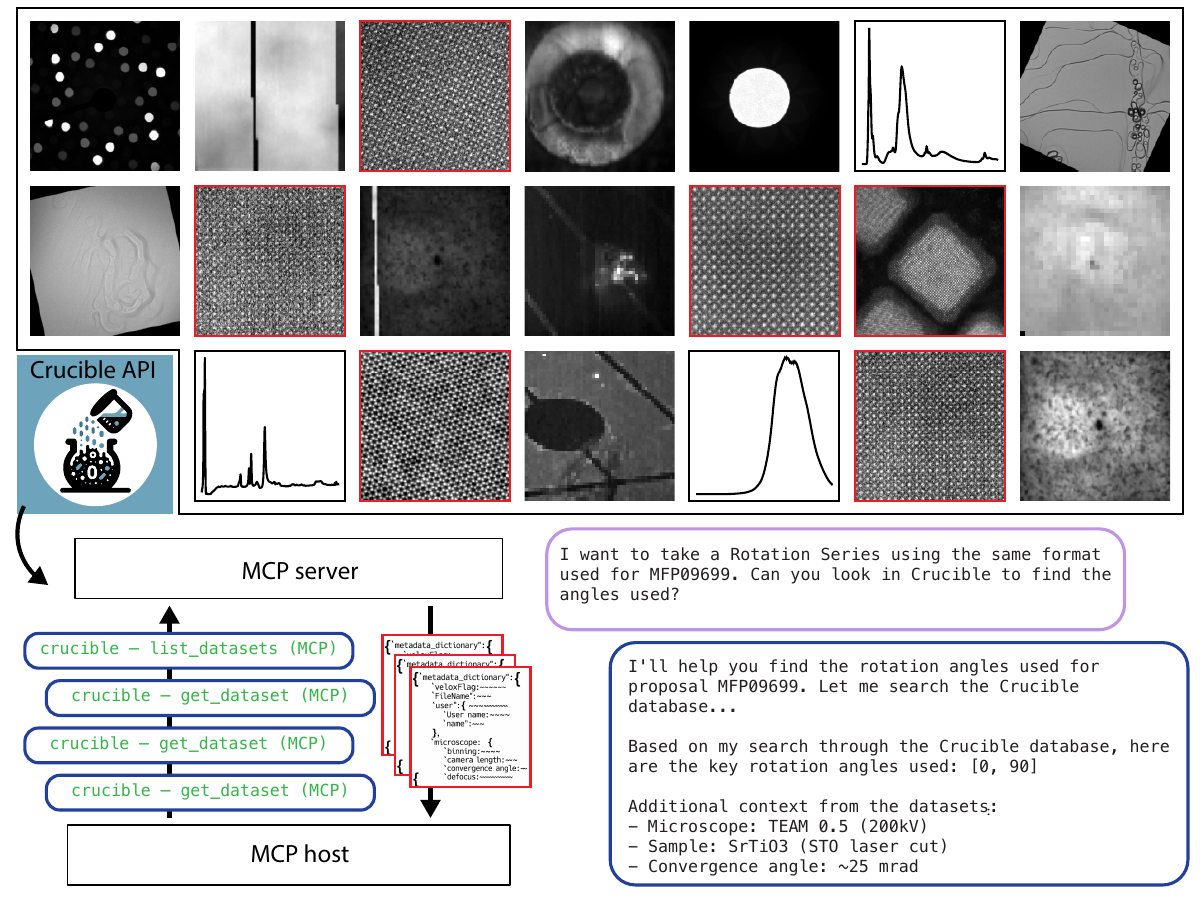}
  \caption{TEM Agent allows seamless integration of our Crucible database into an experiment. Crucible contains experimental data and metadata that is accessible through an API. \texttt{TEM Agent} can interpret user prompts to identify and apply the filters to find relevant historical datasets and then drieclty apply those settings to the microscope.}
  \label{fig:historical}
\end{figure*}

Implementation of MCP servers with a common agent enables a simple one-step integration between the previously described \textit{tools} for instrument control software and independently developed data management \textit{tools} and \textit{resources}. 
Data management and curation is typically an afterthought of an experiment and either left to the users (i.e. file transfers to personal storage) or post-session ingestion into a shared resource to assist with post-processing.
In our setup, we extend the information available to the LLM by including an MCP server with \textit{tools} for accessing our internally developed experimental data platform called \texttt{Crucible}~\cite{crucible}.
At the Molecular Foundry, data is collected by many users using many different types of instruments associated with a variety of projects. The Crucible platform automatically associates data acquired by Molecular Foundry users with projects based on their proposal numbers. Crucible platform users who are not a part of the Molecular Foundry may still contribute data to the platform by creating their own projects through the API. Users can then associate their datasets with existing project IDs to enable downstream filtering and access control. 
Prior to MCP development, access to Crucible metadata was primarily available via web applications with limited functionality, particularly in the case of exploring multiple datasets in aggregate or for searches that required complex joins. 
Using our \texttt{Crucible} MCP server, the agent can chain together API requests to query and summarize information about historically relevant datasets based on user provided criteria to inform the LLM and the user about potentially useful criteria for their current experiment.
Using the available data platform API endpoints and Python client functions, we demonstrate the ability to quickly provide information to the current user about the types of datasets they and others have taken in the past; filter past datasets for experimentally relevant results based on current settings, user information, or project details; and apply this information to guide the user's current experimental set up or microscope settings. 

In this example, we ask the agent to provide details about the experimental parameters typically used by this project for STEM rotation series images as shown in Figure \ref{fig:historical} and Supplementary Movie \ref{si:movie3}. 
Rotation series are often collected for post-processing to remove scan artifacts~\cite{Sang2014-gg}.
The agent executes the appropriate MCP task to query the data platform for the rotation series datasets filtered on project name, collects the scientific metadata entry associated with each dataset ID, and summarizes the resulting information into a table of recommended settings based on other available metadata such as notes, keywords, or sample descriptions.
The agent then applies the settings to the microscope and acquires the data using a to do list similarly to the previous tomography example.
The same information could be collected by the agent by providing the data platform as an MCP \textit{resource} or by directly connecting the database and allowing the agent to compose SQL queries. For the purpose of this project, leveraging the existing Python client in combination with an MCP server allows for computationally cheaper and more restrictive, pre-defined database interactions, while enforcing authentication and authorization to the platform at the user level. 

\subsection{Integrating multiple metadata sources and enhancing advanced experiments}
Another strength of the MCP technology that we observed and leveraged in our implementation is the ability to draw information from related but disparate data sources in a centralized, accessible manner. 
In addition to the \texttt{Crucible} data platform for metadata and data cataloging, the Molecular Foundry leverages another platform, \texttt{Distiller}, for managing 4D-STEM dataset metadata acquired by the 4D Camera~\cite{welborn2025streaming}. Connecting both platforms to our framework through independent MCP servers enables \texttt{TEM Agent} to access information from either platform, without the user having to know or specify which database they need to query for their specific experiment. In this final use case, we demonstrate how we can leverage historical metadata from the \texttt{Distiller} platform,  
with specific \textit{tools} to enable complex workflows that would be challenging to implement without past experience. 

\begin{figure*}[ht]
\centering
\includegraphics[width = 0.95\textwidth]{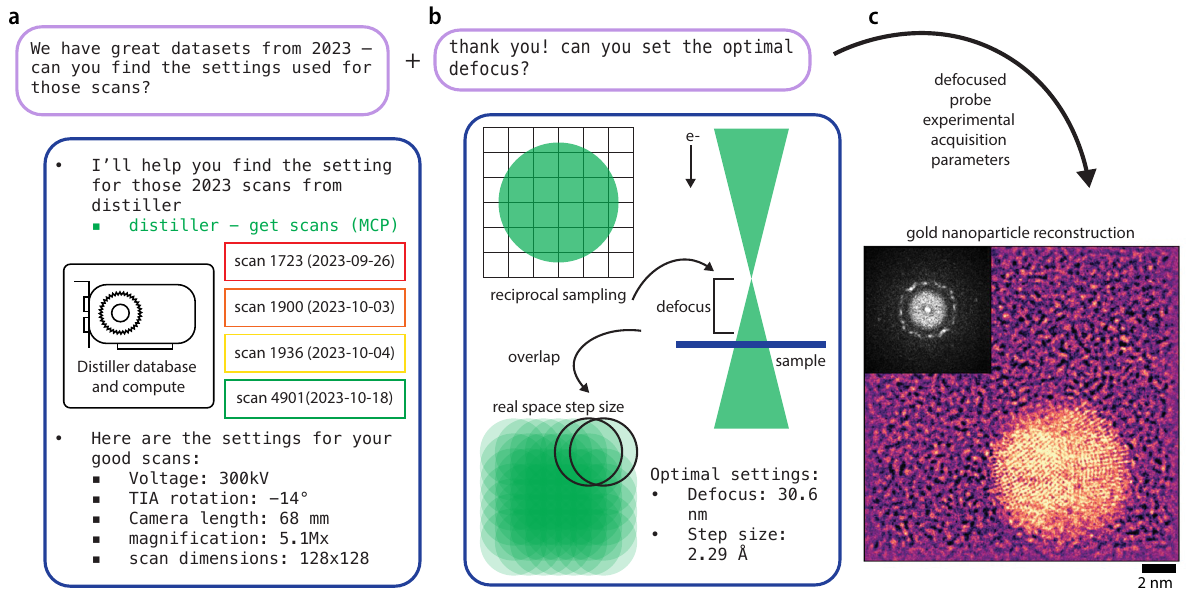}
  \caption{(a) \texttt{TEM agent} queries metadata from \texttt{Distiller}, summarizes the information for the user, (b) and uses the ptychography \textit{tool} to optimize data acquisition conditions. (c) Our framework results in a high-quality final reconstruction of the experimental data (completed offline).}
  \label{fig:ptycho}
\end{figure*}

Iterative electron ptychography and parallax, for example, are STEM phase retrieval methods used for dose-efficient imaging~\cite{yu2025dose, varnavides2023iterative}. 
These experiments are often performed in an unconventional geometry with a defocused probe~\cite{varnavides2023iterative, chen2024imaging}, which is unfamiliar for users who do not have experience with this approach. 
\texttt{TEM Agent} can help facilitate these experiments using information from successful past experiments as well as \textit{tools} designed to improve data quality. 
We began by pointing the LLM to successful scans from our past work~\cite{ribet2024uncovering, pattison2025beacon}, and using notes from \texttt{Distiller}, the LLM guided us towards typical experimental parameters (Fig.~\ref{fig:ptycho}a).
While some metadata is automatically collected during routine microscope operation, other calibrations are only recorded through inconsistent, manually-written user notes in \texttt{Distiller}. 
The ability to search, interpret, and aggregate this varying metadata into specific executable-parameters is a unique benefit of using an LLM in our framework. 

Further, one of the challenges in collecting ptychography and parallax data is optimizing defocus and step-size. 
The success of reconstructions for these experiments is predicated on having sufficient overlap between adjacent scan points in real-space as well as having a reasonable probe size based on the reciprocal sampling (camera length) set on the microscope~\cite{varnavides2025relaxing, varnavides2023iterative}.
These two important parameters are controlled by the STEM probe step size in real space and the STEM camera length, both of which are stored in the \texttt{Distiller} database.
While an experienced microscopist can estimate these parameters during an experiment, this is challenging for new operators. 
Therefore, we developed a \textit{tool} to calculate optimal parameters based on user-tunable inputs and live calibrations on the microscope, which the LLM accessed through the MCP server (Fig.~\ref{fig:ptycho}b).
Ultimately, this data led to a high-quality parallax reconstruction of a gold nanoparticle (Fig.~\ref{fig:ptycho}c).
The experiment is shown in Supplementary Movie \ref{si:movie4}.

It is important to note that the LLM's ability to dynamically interact with the microscope through software and fine-tune parameters allows for more precise data-acquisition than the course-grained control knobs traditionally offered to users. 
The \texttt{TEM Agent} added complete metadata notes to \texttt{Distiller}, facilitating improved data curation and future experiments.
Ultimately, the ability to extract information from large resource libraries and apply this information to the current experiment with help from dedicated \textit{tools} has given our \texttt{TEM Agent} the ability to provide automated data acquisition and support for new users, even on complex workflows.

\section{Discussion} 
\label{sec:discussion}

The development of our \texttt{TEM Agent} framework was driven by our experience at a user facility, where researchers complete short-term projects composed of many different kinds of experiments. 
The diversity of samples and experiments makes it challenging to develop automation workflows. 
However, the combination of MCP servers and an LLM allows for on-the-fly automation and quick implementation of new experimental routines with minimal user training. 
Moreover, this approach removes some of the domain-specific knowledge required to operate microscopes. 
Even though most TEMs are commercial, each microscope has a unique suite of scientific software to facilitate connections to microscope accessories manufactured by different vendors. 
The \texttt{TEM Agent} framework enables expert microscopists to encode their domain knowledge such that it can benefit more users without requiring direct supervision.
It also allows microscope managers to expose different levels of microscope functionality to different users depending on their needs and/or experience.

The LLM itself played a central role in the execution of the experiments described here.
Our use of MCP servers enabled us to leverage the powerful language processing and summarization capabilities of a commercial LLM, both by parsing user input into an experimental plan of technical tasks as well as by summarizing task output and results into meaningful summaries. 
For example, in the tomography experiment, \texttt{TEM Agent} quickly created a to-do list of low-level steps to execute a tomography experiment without any specific training. 
Instead, our framework relied on prompt engineering, \textit{tools}, \textit{resources}, and connections to large libraries of historical data. 

It is noteworthy that the relatively quick success of the MCP integration into our microscope was predicated upon past work to develop low-level Python code and collect metadata as efficiently as possible, which will inform future research efforts.
Specifically, there are significant challenges in coupling hardware and software, especially for advanced tools such as transmission electron microscopes that are made up of sub-components manufactured by a variety of vendors. 
Typical operation involves using different pieces of vendor-specific software that connects to individual sub-components of the microscope, such as an aberration corrector or detector. 
Writing specific code to develop an API to facilitate useful and safe use of these disparate microscope pieces is challenging and time consuming, especially as manual care is needed to port basic functionality to vendor-specific tools.
However, the benefits of writing this low-level code is enormous. 
We have observed this in our automation research, and this value has been demonstrated as well through our \texttt{TEM Agent} approach.
The MCP layer is relatively straightforward to design once there is already sufficient infrastructure in place to facilitate basic microscope operation and could become a standard in the community to interface many different pieces of equipment together.

Similarly, our existing metadata libraries were invaluable in the work here. 
Metadata is an important component of any scientific experiment, but one of the challenges in writing data analysis software across disciplines is the need to develop flexible libraries that support and use the various forms of metadata in data analysis. 
Recognizing the importance of storing this information, we developed support infrastructure for our data analysis, through both the \texttt{Crucible} and \texttt{Distiller} platforms.
The metadata our users record is helpful for their own research, but it would be laborious for a human to parse through all the historical metadata acquired by our microscopes available in the \texttt{Crucible} and \texttt{Distiller} platforms. 
This is relatively simple for LLMs, and our \texttt{TEM Agent} significantly benefits from past efforts to record metadata in a machine readable form. 
Ultimately, this project underscores the value of writing low-level code and storing metadata.

Because our \texttt{TEM Agent} framework does not rely on electron microscopy-specific code development, this approach could be implemented with other instruments. 
Given the similarity between STEM and other scanning probe tools, other scanning microscopes were an obvious choice to extend this framework. This is now possible because many scientific instruments in other domains have APIs that allow for automated control. Such APIs can be wrapped in an MCP server to quickly achieve similar results. Examples include Oxford Asylum Ergo software for atomic force microscopy, Nanonis scanning tunneling microscopy software, and the ScopeFoundry \cite{durham_scopefoundry_2018} instrumentation control platform. While APIs for these software systems are substantially different from an organizational and protocol perspective, a direct mapping from API function calls to MCP \textit{tools} is straightforward and requires minimal software development. The flexibility of the LLM to organize MCP \textit{tool} calls and chain tasks reduces the need to have similar organizational structure across these very different APIs.

Our \texttt{TEM Agent} framework is text-based, and we did not include image processing capabilities.
There are a variety of reasons for this. 
We had initially explored the possibility of returning the entire STEM image to the LLM, but this caused several problems. 
It very quickly increased the number of tokens in the LLM ``context" leading to reduced efficiency and increased computation (costs). 
Further, we learned that the LLM is best suited to interpreting text-based results rather than complex experimental STEM images of a specific unique sample.
In cases where we wanted the LLM to have information about the acquired image in its context, the necessary image processing calculations were added as an MCP \textit{tool} and included in the response (e.g. image intensity standard deviation). 
A future LLM or MCP \textit{tool} could be developed that is capable of image segmentation and labeling, but this would require expensive retraining leading to a local bespoke model, which is beyond the scope of this paper. 
Still, development of a STEM-specific AI model with the ability to interpret such images would greatly benefit the community.

Another limitation of the current implementation is that the workflow determined by the LLM is not deterministic.
On repeated attempts of the same prompt, the LLM suggested different sequences of actions and in many cases seemed to become more creative with repeated attempts. 
Future directions to address this observation and promote more reliable behavior could include leveraging memory, providing one-shot examples, or creating a feedback loop of scored interactions to help the agent learn the desired behavior.
Inclusion of common procedures and sequences in the memory file would be a very simple start to addressing this problem. 
More complex solutions could involve the creation of a more formal Knowledge Base with information about common workflows, instrument manuals, publications with details about past experiments, or increased access to our data and metadata platforms. 
Additionally, prompt and response pairs could be recorded over time as the system is used and responses could be scored, either manually or by an evaluation agent, and then returned to the Knowledge Base to inform future responses.

Lastly, the fact that the LLM interacts with the MCP layer rather writing code to control the microscope or interface with the metadata libraries safeguarded against potential mistakes in LLM-written code seriously harming either system.
Each MCP tool was designed such that even its misapplication by the LLM would not damage the microscope or corrupt the metadata library.
Mitigating possible issues like this is important to consider given the cost of the microscope and the large potential user base at the Foundry.
However, there are promising routes forward towards more agentic frameworks including open source software such as the Alpha Berkeley project \cite{Hellert2025-pr}. 
While Claude Code has provided a straightforward, simple, and distributed solution for connecting the underlying LLM with the MCP servers, evaluation of other interfaces or agentic platforms could provide a more intuitive entry point for less technical users. 
Facilitating fully agentic operation will also require hardware and software development to ensure safe operation of the microscope. 

\section{Conclusion}

We have demonstrated the capabilities of our \texttt{TEM Agent} framework in the context of simplifying the operation of a complex TEM.
We show how the combination of a general LLM with domain-specific MCP servers can improve the user experience by automating tedious tasks while also enabling more involved routines such as tomography.
We also showed the utility of incorporating metadata from past experiments and experiment specific \textit{tools}, such as for calculating ptychography defocus.
Agentic operation of a TEM using \texttt{TEM Agent} thus could lead to improved data quality and throughput with better scientific outcomes.

Especially in the context of a user facility, \texttt{TEM Agent} is able to bridge the gap in expertise by providing the ability to accomplish complex tasks and summarize results using only natural language familiar to most users.
We see this type of operation as a highly flexible way to interact with a TEM, allowing more exploration of microscope capabilities and parameters beyond what is typically provided in standard operating procedures.
The automation demonstrated here is certainly useful, but traditional software programs developed to accomplish a dedicated, specific task would likely outperform any agent in the near term.

Moreover, we did not incorporate any microscopy-specific training on images or text, so the LLM has a limited fundamental understanding of microscopy. 
We instead used \textit{tools} and \textit{resources} to help guide \texttt{TEM Agent} microscope operation, but more training would facilitate advanced, on-the-fly analysis.
Moreover, the framework only knows about the hardware components that are incorporated, so the full microscope is not accessible.
This approach does not make sense for all cases: many experiments require tasks that would be faster to execute than optimize even with an LLM, while others require advanced analysis that would only be possible through specific machine learning training or human feedback.
At the moment, having a human-in-the-loop is still an important part of this framework, and ultimately, there is no replacement for an experienced human operator, who requires time operating the microscope manually to gain experience.

Overall, our \texttt{TEM Agent} experiments highlighted the benefits of the MCP framework for ease of integrating LLMs into an experimental, hardware driven ecosystem, and the impact that it can have on microscope usage. 
However, it is important to note that this framework relies heavily on existing LLM capabilities. 
A variety of societal factors far outside of the scope of this paper have driven the rise of LLMs in the past few years, which come with their own set of ethical considerations, including cost, accessibility, environmental impact, and privacy. 
Future LLM technological development and policies will drive innovation in our context as well. 

\section{Methods}

\subsection{Microscope server}
\label{sec:microscope_server}
Many different operations of the TEAM 0.5 electron microscope are exposed to the user on a Windows COM interface by the vendor (Thermo Fischer, formerly FEI).
We previously made available many of these functions to the remote Microscope Support PC through a custom zeroMQ server~\cite{pattison2023advanced}.
The server is able to query and update states of the microscope and return the information to a remote client over a local area network. Our server communicates with the TEMScripting microscope interface, the TIA data acquisition interface, and the aberration corrector.
The available read/write microscope parameters include defocus, stage position, goniometer tilt, beam tilt, and beam blanking. 
The server also implements several common sets of microscope actions such as auto-focusing by Bayesian optimization and image acquisition. 
These actions take simple inputs that then implement the many required tasks to accomplish the final goal. 
For example, STEM image acquisition requires inputs such as detector choice, number of fast-scan pixels, number of slow-scan pixels, dwell time, continuous/single, etc. 
We expose only one function that reliably sets up the STEM scanning server properly each time with minimal inputs.
Image data is automatically saved to disk with a unique filename, and only metadata (such as the image intensity minimum, maximum, standard deviation and saved file name) are returned to the LLM.
We purposely chose not to return image data to the LLM as described in the \hyperref[sec:discussion]{Discussion} section.

\subsection{Detector server}
\label{sec:detector_server}
The TEAM 0.5 microscope is equipped with the 4D Camera and Digiscan scanning system controlled by the Gatan Digital Micrograph (DM) software installed on a separate vendor computer. We implemented a zeroMQ server that receives commands to execute custom templated scripts in the Digital Micrograph software using system calls. More information is available elsewhere.\cite{pattison2023advanced}

\subsection{Distiller}

The \texttt{Distiller} platform allows users at the TEAM 0.5 microscope to monitor and interact with data generated by the 4D Camera. It allows real-time monitoring, launches jobs on the NERSC high performance computer (HPC), handles metadata, and integrates with Jupyter notebooks. The frontend is a web application based on a React framework and the backend utilizes FastAPI, Kafka, and Microservices to provide integration with the NERSC Superfacility API~\cite{Enders2020-ez}. The web application shows information about all scans acquired with the 4D Camera. Each dataset is tagged with a unique scan identification number and is associated with the simultaneously acquired HAADF-STEM image, microscope metadata, and user-entered notes. All data is stored in a database to assist in post-processing. The platform also allows the user to launch jobs on the NERSC supercomputer to start a file copy process for data processing and to instantiate a streaming session for direct-to-RAM data processing~\cite{Welborn2025-uz}. Our MCP server utilized the FastAPI endpoints to download metadata and save notes in the system. 

\subsection{Crucible}
The \texttt{Crucible} Data Platform is a FastAPI application with a PostgreSQL database backend deployed on the Google Cloud Platform. The platform has been developed in accordance with FAIR (findable, accessible, interoperable, reusable) data principles, to provide a centralized, searchable, and accessible storage solution for experimental data acquired on instruments at the Molecular Foundry. Small datasets of a wide variety of formats can be uploaded to the platform via the API and metadata is extracted into a standardized set of fields that are then available both as JSON files in s3 object storage and SQL records in the database. Additionally, scientific metadata about the experimental setup that may not be preserved across measurement types or instruments is captured as nested JSON. This information can then be queried or sent to downstream services such as our data catalog, hosted using the open source software SciCat \cite{Malviya-Thakur2023-xd}. To execute data management tasks such as extracting metadata, sending the metadata to downstream services, or transferring the data to other storage locations, API requests can be made to specific endpoints. These endpoints will then update the backend database to reflect the task request and publish a message to a RabbitMQ server. The RabbitMQ server will then route queued messages to the appropriate downstream consumers for processing.  

\subsection{MCP server set-up}
Using the MCP, four modules were integrated with Claude Code to provide a unified interface for communicating with the microscope server, communicating with the 4D camera, accessing the \texttt{Crucible} data platform, and accessing the Distiller API. The FastMCP Python package was used to add an MCP layer to each module.  Modular development of each MCP server not only allows tools to be added and developed by functionality, but also allows distributed deployment of the various servers, while still preserving the unified interface.  MCP servers responsible for communicating with the 4D Camera, \texttt{Distiller}, and TEAM 0.5 microscope are deployed on Microscope Support PC using http transport. The \texttt{Crucible} MCP server was run locally on the user's laptop and communicates using the stdio transport with Claude Code. This design offers the advantage of streamlining credentials and networking configuration between all of the various modules.  Credentials for interacting with the instrument control interfaces are never exposed to the user and are only stored on the local instrument control computers. Conversely, credentials for interacting with user specific data and metadata are managed at the client level.  A more thorough evaluation of agentic tools and interfaces for unifying the MCP modules is a future direction that we would like to explore.

\subsection{Data acquisition}

Data were acquired on the TEAM 0.5 which is a double-aberration-corrected transmission electron microscope hosted by the National Center for electron Microscopy in the the Molecular Foundry at Lawrence Berkeley National Laboratory. 
In all experiments, only the LLM made changes to the microscope.
The parallax data were collected with the 4D Camera running at 87,000 frames per second~\cite{Ercius2024-ed}. 
The microscope was operated at 300 kV in all experiments.
The sample that was studied was gold nanoparticles supported on a ultra-thin carbon grid.

The focus optimization experiment described in the Section \ref{sec:simple} was acquired with a convergence angle of 25 mrad and approximately 20 pA of beam current. The focus was changed in steps of 20 nm. Image settings were 512 by 512 pixels, a probe step size of 0.17 nm, and a field of view of 86.87 nm. 

Tomography data were acquired from 0$^\circ$ to 20$^\circ$ in 5$^\circ$ steps. 
The stage eucentric height was manually set before data acquisition.
Images with 512 by 512 pixels were acquired with a dwell time of 10$\mu$s, a convergence angle of 30 mrad, and approximately 35 pA of beam current. The tilt series image acquisitions used a 10.9 nm probe step size. After each tilt, the central region of the image was focused using the BEACON Bayesian optimization routine. The central 256 by 512 region was used. 

Parallax/ptychography data was acquired with a camera length of 92 mm (68 mm indicated by the user interface). 
Using 85\% overlap and an estimated reciprocal sampling of 0.0218 {\AA}$^{-1}$, the optimal defocus of 30.6 nm and 2.2 {\AA} step size were selected. 
A 128 by 128 scan was acquired and four frames were acquired at each probe position for a total dwell time of 44 $\mu$s. The applied dose is approximately $2\times10^3$ e$^-$/\AA$^2$. The convergence angle was 20 mrad.
The parallax reconstruction was performed in \texttt{py4DSTEM}~\cite{savitzky2021py4dstem, varnavides2023iterative}.

The ptychography \textit{tool} to determine the optimal defocus was built using the following logic.
The probe cropping box for a reconstruction is the inverse of the reciprocal sampling. 
The probe diameter should be approximately 1/3 of the box size. 
Using a known convergence angle, the defocus to give a known probe size at the sample can be estimated. 
From here, the step size can be calculated with an optimal overlap with the default set at 85\%.


\section*{Code Availability}

Open-source code for the control of TEAM 0.5 including the microscopy, detectors, and Distiller is freely available at \href{https://github.com/foundry-mcp/team05-mcp-server}{https://github.com/foundry-mcp/team05-mcp-server}. Open-source code to retrieve data from Crucible is freely available at \href{https://github.com/foundry-mcp/mcp-crucible}{https://github.com/foundry-mcp/mcp-crucible}.

\bibliography{sn-bibliography}

@ARTICLE{Sang2014-gg,
  title     = "Revolving scanning transmission electron microscopy: correcting
               sample drift distortion without prior knowledge",
  author    = "Sang, Xiahan and LeBeau, James M",
  journal   = "Ultramicroscopy",
  publisher = "Elsevier BV",
  volume    =  138,
  pages     = "28--35",
  month     =  mar,
  year      =  2014,
  doi       = "10.1016/j.ultramic.2013.12.004"
}

@ARTICLE{Lyu2023-xg,
  title    = "Electron Microscopy Studies of Soft Nanomaterials",
  author   = "Lyu, Zhiheng and Yao, Lehan and Chen, Wenxiang and Kalutantirige,
              Falon C and Chen, Qian",
  journal  = "Chemical reviews",
  volume   =  123,
  number   =  7,
  pages    = "4051--4145",
  month    =  apr,
  year     =  2023,
  doi      = "10.1021/acs.chemrev.2c00461"
}

@ARTICLE{Ophus2023-pk,
  title     = "Quantitative Scanning Transmission Electron Microscopy for
               Materials Science: Imaging, Diffraction, Spectroscopy, and
               Tomography",
  author    = "Ophus, Colin",
  journal   = "Annual review of materials research",
  publisher = "Annual Reviews",
  month     =  jul,
  year      =  2023,
  doi       = "10.1146/annurev-matsci-080921-092646"
}

@INCOLLECTION{Welborn2025-uz,
  title     = "Accelerating time-to-science by streaming detector data directly
               into Perlmutter compute nodes",
  author    = "Welborn, Samuel S and Harris, Chris and Ercius, Peter and Bard,
               Deborah J and Enders, Bjoern",
  booktitle = "Lecture Notes in Computer Science",
  publisher = "Springer Nature Switzerland",
  address   = "Cham",
  pages     = "243--256",
  series    = "Lecture notes in computer science",
  year      =  2025,
  doi       = "10.1007/978-3-031-73716-9\_17"
}

@INPROCEEDINGS{Enders2020-ez,
  title     = "Cross-facility science with the Superfacility Project at {LBNL}",
  author    = "Enders, Bjoern and Bard, Debbie and Snavely, Cory and Gerhardt,
               Lisa and Lee, Jason and Totzke, Becci and Antypas, Katie and
               Byna, Suren and Cheema, Ravi and Cholia, Shreyas and Day, Mark
               and Gaur, Aditi and Greiner, Annette and Groves, Taylor and
               Kiran, Mariam and Koziol, Quincey and Rowland, Kelly and Samuel,
               Chris and Selvarajan, Ashwin and Sim, Alex and Skinner, David and
               Thomas, Rollin and Torok, Gabor",
  booktitle = "2020 IEEE/ACM 2nd Annual Workshop on Extreme-scale
               Experiment-in-the-Loop Computing (XLOOP)",
  publisher = "IEEE",
  pages     = "1--7",
  month     =  nov,
  year      =  2020,
  doi       = "10.1109/xloop51963.2020.00006",
  address   = "New York, NY"
}

@ARTICLE{Yager2024-ln,
  title     = "Towards a science exocortex",
  author    = "Yager, Kevin G",
  journal   = "Digital discovery",
  publisher = "Royal Society of Chemistry (RSC)",
  volume    =  3,
  number    =  10,
  pages     = "1933--1957",
  month     =  oct,
  year      =  2024,
  doi       = "10.1039/d4dd00178h",
  issn      = "2635-098X",
  language  = "en"
}

@ARTICLE{Hellert2025-pr,
  title         = "Alpha Berkeley: A scalable framework for the orchestration of
                   agentic systems",
  author        = "Hellert, Thorsten and Montenegro, João and Sulc, Antonin",
  journal       = "arXiv [cs.MA]",
  month         =  aug,
  year          =  2025,
  archivePrefix = "arXiv",
  primaryClass  = "cs.MA",
  eprint        = "2508.15066",
  doi           = "10.48550/arXiv.2508.15066"
}

@ARTICLE{Malviya-Thakur2023-xd,
  title         = "{SciCat}: A curated dataset of scientific software
                   repositories",
  author        = "Malviya-Thakur, Addi and Milewicz, Reed and Paganini, Lavinia
                   and Mahmoud, Ahmed Samir Imam and Mockus, Audris",
  journal       = "arXiv [cs.SE]",
  month         =  dec,
  year          =  2023,
  archivePrefix = "arXiv",
  primaryClass  = "cs.SE",
  eprint        = "2312.06382",
  doi           = "10.48550/arXiv.2312.06382"
}

@ARTICLE{Mastronarde2005-zo,
  title    = "Automated electron microscope tomography using robust prediction
              of specimen movements",
  author   = "Mastronarde, David N",
  journal  = "Journal of structural biology",
  volume   =  152,
  number   =  1,
  pages    = "36--51",
  month    =  oct,
  year     =  2005,
  doi      = "10.1016/j.jsb.2005.07.007"
}

@ARTICLE{Mastronarde2017-dm,
  title     = "Automated tilt series alignment and tomographic reconstruction in
               {IMOD}",
  author    = "Mastronarde, David N and Held, Susannah R",
  journal   = "Journal of structural biology",
  publisher = "Academic Press",
  volume    =  197,
  number    =  2,
  pages     = "102--113",
  month     =  feb,
  year      =  2017,
  doi       = "10.1016/j.jsb.2016.07.011"
}

@ARTICLE{Ercius2015-un,
  title     = "Electron tomography: A three-dimensional analytic tool for hard
               and soft materials research",
  author    = "Ercius, Peter and Alaidi, Osama and Rames, Matthew J and Ren,
               Gang",
  journal   = "Advanced materials (Deerfield Beach, Fla.)",
  publisher = "Wiley",
  volume    =  27,
  number    =  38,
  pages     = "5638--5663",
  month     =  oct,
  year      =  2015,
  doi       = "10.1002/adma.201501015"
}

@ARTICLE{Tabor2018-fk,
  title    = "Accelerating the discovery of materials for clean energy in the
              era of smart automation",
  author   = "Tabor, Daniel P and Roch, Loïc M and Saikin, Semion K and
              Kreisbeck, Christoph and Sheberla, Dennis and Montoya, Joseph H
              and Dwaraknath, Shyam and Aykol, Muratahan and Ortiz, Carlos and
              Tribukait, Hermann and Amador-Bedolla, Carlos and Brabec,
              Christoph J and Maruyama, Benji and Persson, Kristin A and
              Aspuru-Guzik, Alán",
  journal  = "Nature Reviews Materials",
  volume   =  3,
  number   =  5,
  pages    = "5--20",
  month    =  may,
  year     =  2018,
  doi      = "10.1038/s41578-018-0005-z"
}

@ARTICLE{Kalinin2022-ga,
  title     = "Machine learning in scanning transmission electron microscopy",
  author    = "Kalinin, Sergei V and Ophus, Colin and Voyles, Paul M and Erni,
               Rolf and Kepaptsoglou, Demie and Grillo, Vincenzo and Lupini,
               Andrew R and Oxley, Mark P and Schwenker, Eric and Chan, Maria K
               Y and Etheridge, Joanne and Li, Xiang and Han, Grace G D and
               Ziatdinov, Maxim and Shibata, Naoya and Pennycook, Stephen J",
  journal   = "Nature reviews. Methods primers",
  publisher = "Springer Science and Business Media LLC",
  volume    =  2,
  number    =  1,
  pages     = "1--28",
  month     =  mar,
  year      =  2022,
  doi       = "10.1038/s43586-022-00095-w"
}

@article{yin2024pear,
  title={PEAR: a robust and flexible automation framework for ptychography enabled by multiple large language model agents},
  author={Yin, Xiangyu and Shi, Chuqiao and Han, Yimo and Jiang, Yi},
  journal={arXiv preprint arXiv:2410.09034},
  year={2024},
doi="10.48550/arXiv.2410.09034"
}

@article{jiang2025empowering,
  title={Empowering Electron Ptychography with Generative Artificial Intelligence and Agentic Workflows},
  author={Jiang, Yi and Yin, Xiangyu and Shi, Chuqiao and Fein-Ashley, Benjamin and Shao, Yu-Tsun and Han, Yimo},
  journal={Microscopy and Microanalysis},
  volume={31},
  number={Supplement\_1},
  pages={ozaf048--1061},
  year={2025},
  publisher={Oxford University Press US},
doi="10.1093/mam/ozaf048.1061"
}

@article{varnavides2023iterative,
  title={Iterative phase retrieval algorithms for scanning transmission electron microscopy},
  author={Varnavides, Georgios and Ribet, Stephanie M and Zeltmann, Steven E and Yu, Yue and Savitzky, Benjamin H and Byrne, Dana O and Allen, Frances I and Dravid, Vinayak P and Scott, Mary C and Ophus, Colin},
  journal={arXiv preprint arXiv:2309.05250},
  year={2023}
}

@article{savitzky2021py4dstem,
    author = {Savitzky, Benjamin H and Zeltmann, Steven E and Hughes, Lauren A and Brown, Hamish G and Zhao, Shiteng and Pelz, Philipp M and Pekin, Thomas C and Barnard, Edward S and Donohue, Jennifer and Rangel DaCosta, Luis and Kennedy, Ellis and Xie, Yujun and Janish, Matthew T and Schneider, Matthew M and Herring, Patrick and Gopal, Chirranjeevi and Anapolsky, Abraham and Dhall, Rohan and Bustillo, Karen C and Ercius, Peter and Scott, Mary C and Ciston, Jim and Minor, Andrew M and Ophus, Colin},
    title = {py4DSTEM: A Software Package for Four-Dimensional Scanning Transmission Electron Microscopy Data Analysis},
    journal = {Microscopy and Microanalysis},
    volume = {27},
    number = {4},
    pages = {712-743},
    year = {2021},
    month = {08},
    issn = {1431-9276},
}

@article{welborn2025streaming,
  title={Streaming Large-Scale Microscopy Data to a Supercomputing Facility},
  author={Welborn, Samuel S and Harris, Chris and Ribet, Stephanie M and Varnavides, Georgios and Ophus, Colin and Enders, Bjoern and Ercius, Peter},
  journal={Microscopy and Microanalysis},
  volume={31},
  number={1},
  pages={ozae109},
  year={2025},
  publisher={Oxford University Press US},
doi="10.1093/mam/ozae109"
}

@ARTICLE{Ercius2024-ed,
  title     = "The {4D} Camera: An 87 {kHz} direct electron detector for
               scanning/transmission electron microscopy",
  author    = "Ercius, Peter and Johnson, Ian J and Pelz, Philipp and Savitzky,
               Benjamin H and Hughes, Lauren and Brown, Hamish G and Zeltmann,
               Steven E and Hsu, Shang-Lin and Pedroso, Cassio C S and Cohen,
               Bruce E and Ramesh, Ramamoorthy and Paul, David and Joseph, John
               M and Stezelberger, Thorsten and Czarnik, Cory and Lent, Matthew
               and Fong, Erin and Ciston, Jim and Scott, Mary C and Ophus, Colin
               and Minor, Andrew M and Denes, Peter",
  journal   = "Microscopy and microanalysis",
  publisher = "Oxford University Press",
  pages     = "ozae086",
  month     =  sep,
  year      =  2024,
  doi       = "10.1093/mam/ozae086",
  pmid      =  39298134,
  issn      = "1431-9276,1435-8115",
  language  = "en"
}

@article{pattison2025beacon,
  title={BEACON—automated aberration correction for scanning transmission electron microscopy using Bayesian optimization},
  author={Pattison, Alexander J and Ribet, Stephanie M and Noack, Marcus M and Varnavides, Georgios and Park, Kunwoo and Kirkland, Earl J and Park, Jungwon and Ophus, Colin and Ercius, Peter},
  journal={npj Computational Materials},
  volume={11},
  number={1},
  pages={274},
  year={2025},
  publisher={Nature Publishing Group UK London},
doi="10.1038/s41524-025-01766-4"
}

@article{kirkland2018fine,
  title={Fine tuning an aberration corrected ADF-STEM},
  author={Kirkland, Earl J},
  journal={Ultramicroscopy},
  volume={186},
  pages={62--65},
  year={2018},
  publisher={Elsevier},
doi="10.1016/j.ultramic.2017.12.002"
}

@article{ishikawa2021automated,
  title={Automated geometric aberration correction for large-angle illumination STEM},
  author={Ishikawa, Ryo and Tanaka, Riku and Morishita, Shigeyuki and Kohno, Yuji and Sawada, Hidetaka and Sasaki, Takuya and Ichikawa, Masanari and Hasegawa, Masashi and Shibata, Naoya and Ikuhara, Yuichi},
  journal={Ultramicroscopy},
  volume={222},
  pages={113215},
  year={2021},
  publisher={Elsevier},
doi="10.1016/j.ultramic.2021.113215"
}

@ARTICLE{kalinin2023machine,
  title     = "Machine learning for automated experimentation in scanning
               transmission electron microscopy",
  author    = "Kalinin, Sergei V and Mukherjee, Debangshu and Roccapriore, Kevin
               and Blaiszik, Benjamin J and Ghosh, Ayana and Ziatdinov, Maxim A
               and Al-Najjar, Anees and Doty, Christina and Akers, Sarah and
               Rao, Nageswara S and Agar, Joshua C and Spurgeon, Steven R",
  journal   = "npj Computational Materials",
  publisher = "Nature Publishing Group",
  volume    =  9,
  number    =  1,
  pages     = "1--16",
  month     =  dec,
  year      =  2023,
  doi       = "10.1038/s41524-023-01142-0"
}

@ARTICLE{spurgeon2021towards,
  title    = "Towards data-driven next-generation transmission electron
              microscopy",
  author   = "Spurgeon, Steven R and Ophus, Colin and Jones, Lewys and
              Petford-Long, Amanda and Kalinin, Sergei V and Olszta, Matthew J
              and Dunin-Borkowski, Rafal E and Salmon, Norman and Hattar, Khalid
              and Yang, Wei-Chang D and Sharma, Renu and Du, Yingge and
              Chiaramonti, Ann and Zheng, Haimei and Buck, Edgar C and Kovarik,
              Libor and Penn, R Lee and Li, Dongsheng and Zhang, Xin and
              Murayama, Mitsuhiro and Taheri, Mitra L",
  journal  = "Nature materials",
  volume   =  20,
  number   =  3,
  pages    = "274--279",
  month    =  mar,
  year     =  2021,
  doi      = "10.1038/s41563-020-00833-z"
}

@article{mathur2025vision,
  title={VISION: a modular AI assistant for natural human-instrument interaction at scientific user facilities},
  author={Mathur, Shray and van der Vleuten, Noah and Yager, Kevin G and Tsai, Esther HR},
  journal={Machine Learning: Science and Technology},
  volume={6},
  number={2},
  pages={025051},
  year={2025},
  publisher={IOP Publishing},
  doi={10.1088/2632-2153/add9e4}
}

@article{yu2025dose,
  title={Dose-efficient cryo-electron microscopy for thick samples using tilt-corrected scanning transmission electron microscopy},
  author={Yu, Yue and Spoth, Katherine A and Colletta, Michael and Nguyen, Kayla X and Zeltmann, Steven E and Zhang, Xiyue S and Paraan, Mohammadreza and Kopylov, Mykhailo and Dubbeldam, Charlie and Serwas, Daniel and others},
  journal={Nature Methods},
  pages={1--11},
  year={2025},
  publisher={Nature Publishing Group US New York}
}

@article{varnavides2025relaxing,
  title={Relaxing Direct Ptychography Sampling Requirements via Parallax Imaging Insights},
  author={Varnavides, Georgios and Bekkevold, Julie Marie and Ribet, Stephanie M and Scott, Mary C and Jones, Lewys and Ophus, Colin},
  journal={arXiv preprint arXiv:2507.18610},
  year={2025}
}

@article{zimmermann202532,
  title={32 examples of {LLM} applications in materials science and chemistry: towards automation, assistants, agents, and accelerated scientific discovery},
  author={Zimmermann, Yoel and Bazgir, Adib and Al-Feghali, Alexander and Ansari, Mehrad and Bocarsly, Joshua and Brinson, L Catherine and Chiang, Yuan and Circi, Defne and Chiu, Min-Hsueh and Daelman, Nathan and others},
  journal={Machine Learning: Science and Technology},
  year={2025},
doi="10.1088/2632-2153/ae011a"
}

@misc{schlozma_llm_autotem,
  author = {Schloz, Marcel  and Gonzalez, Jose Cojal},
  title = {LLM\_autoTEM},
  year = {2024},
  url = {https://gitlab.com/Schlozma/llm_autotem},
  journal = {GitLab repository}
}

@article{zimmermann2024reflections,
  title={Reflections from the 2024 large language model (llm) hackathon for applications in materials science and chemistry},
  author={Zimmermann, Yoel and Bazgir, Adib and Afzal, Zartashia and Agbere, Fariha and Ai, Qianxiang and Alampara, Nawaf and Al-Feghali, Alexander and Ansari, Mehrad and Antypov, Dmytro and Aswad, Amro and others},
  journal={arXiv preprint arXiv:2411.15221},
  year={2024},
doi="10.48550/arXiv.2411.15221"
}

@article{yang2025automat,
  title={AutoMat: Enabling Automated Crystal Structure Reconstruction from Microscopy via Agentic Tool Use},
  author={Yang, Yaotian and Tang, Yiwen and Chen, Yizhe and Chen, Xiao and Qiu, Jiangjie and Xiong, Hao and Yin, Haoyu and Luo, Zhiyao and Zhang, Yifei and Tao, Sijia and others},
  journal={arXiv preprint arXiv:2505.12650},
  year={2025},
doi="10.48550/arXiv.2505.12650"
}

@ARTICLE{li2020robot,
  title     = "Robot-Accelerated Perovskite Investigation and Discovery",
  author    = "Li, Zhi and Najeeb, Mansoor Ani and Alves, Liana and Sherman,
               Alyssa Z and Shekar, Venkateswaran and Cruz Parrilla, Peter and
               Pendleton, Ian M and Wang, Wesley and Nega, Philip W and Zeller,
               Matthias and Schrier, Joshua and Norquist, Alexander J and Chan,
               Emory M",
  journal   = "Chemistry of materials: a publication of the American Chemical
               Society",
  publisher = "American Chemical Society",
  volume    =  32,
  number    =  13,
  pages     = "5650--5663",
  month     =  jul,
  year      =  2020,
  doi       = "10.1021/acs.chemmater.0c01153"
}

@ARTICLE{szymanski2023autonomous,
  title     = "An autonomous laboratory for the accelerated synthesis of novel
               materials",
  author    = "Szymanski, Nathan J and Rendy, Bernardus and Fei, Yuxing and
               Kumar, Rishi E and He, Tanjin and Milsted, David and McDermott,
               Matthew J and Gallant, Max and Cubuk, Ekin Dogus and Merchant,
               Amil and Kim, Haegyeom and Jain, Anubhav and Bartel, Christopher
               J and Persson, Kristin and Zeng, Yan and Ceder, Gerbrand",
  journal   = "Nature",
  publisher = "Nature Publishing Group",
  volume    =  624,
  number    =  7990,
  pages     = "86--91",
  month     =  dec,
  year      =  2023,
  doi       = "10.1038/s41586-023-06734-w"
}

@ARTICLE{stroppa2023stem,
  title     = "From {STEM} to {4D} {STEM}: Ultrafast Diffraction Mapping with a
               Hybrid-Pixel Detector",
  author    = "Stroppa, Daniel G and Meffert, Matthias and Hoermann, Christoph
               and Zambon, Pietro and Bachevskaya, Darya and Remigy, Hervé and
               Schulze-Briese, Clemens and Piazza, Luca",
  journal   = "Microscopy today",
  publisher = "Oxford Academic",
  volume    =  31,
  number    =  2,
  pages     = "10--14",
  month     =  apr,
  year      =  2023,
  doi       = "10.1093/mictod/qaad005"
}

@ARTICLE{jain2013commentary,
  title     = "Commentary: The Materials Project: A materials genome approach to
               accelerating materials innovation",
  author    = "Jain, Anubhav and Ong, Shyue Ping and Hautier, Geoffroy and Chen,
               Wei and Richards, William Davidson and Dacek, Stephen and Cholia,
               Shreyas and Gunter, Dan and Skinner, David and Ceder, Gerbrand
               and Persson, Kristin A",
  journal   = "APL Materials",
  publisher = "American Institute of Physics",
  volume    =  1,
  number    =  1,
  pages     =  011002,
  month     =  jul,
  year      =  2013,
  doi       = "10.1063/1.4812323"
}

@ARTICLE{ophus2019four,
  title     = "Four-dimensional scanning transmission electron microscopy
               ({4D}-{STEM}): From scanning nanodiffraction to ptychography and
               beyond",
  author    = "Ophus, Colin",
  journal   = "Microscopy and microanalysis",
  publisher = "Cambridge University Press (CUP)",
  volume    =  25,
  number    =  3,
  pages     = "563--582",
  month     =  jun,
  year      =  2019,
  doi       = "10.1017/S1431927619000497"
}

@article{mastronarde2005automated,
  title={Automated electron microscope tomography using robust prediction of specimen movements},
  author={Mastronarde, David N},
  journal={Journal of structural biology},
  volume={152},
  number={1},
  pages={36--51},
  year={2005},
  publisher={Elsevier},
  doi="10.1016/j.jsb.2005.07.007"
}

@article{pattison2023advanced,
  title={Advanced techniques in automated high-resolution scanning transmission electron microscopy},
  author={Pattison, Alexander J and Pedroso, Cassio CS and Cohen, Bruce E and Ondry, Justin C and Alivisatos, A Paul and Theis, Wolfgang and Ercius, Peter},
  journal={Nanotechnology},
  volume={35},
  number={1},
  pages={015710},
  year={2023},
  publisher={IOP Publishing},
    doi="10.1088/1361-6528/acf938"
}

@article{olszta2022automated,
  title={An automated scanning transmission electron microscope guided by sparse data analytics},
  author={Olszta, Matthew and Hopkins, Derek and Fiedler, Kevin R and Oostrom, Marjolein and Akers, Sarah and Spurgeon, Steven R},
  journal={Microscopy and Microanalysis},
  volume={28},
  number={5},
  pages={1611--1621},
  year={2022},
  publisher={Cambridge University Press},
doi="10.1017/S1431927622012065"
}

@ARTICLE{byrne2025neutral,
  title    = "Neutral but impactful: Gallium cluster-induced nanopores from
              beam-blanked gallium ion sources",
  author   = "Byrne, Dana O and Ribet, Stephanie M and Bustillo, Karen C and
              Ophus, Colin and Allen, Frances I",
  journal  = "Microscopy and microanalysis",
  volume   =  31,
  number   =  4,
  month    =  jul,
  year     =  2025,
  doi      = "10.1093/mam/ozaf059"
}

@ARTICLE{bustillo2025data,
  title     = "Data management and analysis at the national center for electron
               microscopy",
  author    = "Bustillo, Karen C and Wall, Morgan K and Pattison, Alexander J
               and Ribet, Stephanie M and Barnard, Edward S and Ercius, Peter E",
  journal   = "Microscopy and microanalysis: the official journal of Microscopy
               Society of America, Microbeam Analysis Society, Microscopical
               Society of Canada",
  publisher = "Oxford University Press (OUP)",
  volume    =  31,
  number    = "Supplement\_1",
  pages     = "ozaf048.1160",
  month     =  jul,
  year      =  2025,
  doi       = "10.1093/mam/ozaf048.1160"
}

@article{wilkinson2016fair,
  title={The FAIR Guiding Principles for scientific data management and stewardship},
  author={Wilkinson, Mark D and Dumontier, Michel and Aalbersberg, IJsbrand Jan and Appleton, Gabrielle and Axton, Myles and Baak, Arie and Blomberg, Niklas and Boiten, Jan-Willem and da Silva Santos, Luiz Bonino and Bourne, Philip E and others},
  journal={Scientific data},
  volume={3},
  number={1},
  pages={1--9},
  year={2016},
  publisher={Nature Publishing Group},
doi="10.1038/sdata.2016.18"
}

@article{pratiush2025stem,
  title={STEM Orchestrator: Managing Multi-hardware-component STEM Automation Seamlessly},
  author={Pratiush, Utkarsh and Houston, Austin and Longo, Paolo and Geurts, Remco and Kalinin, Sergei V and Duscher, Gerd},
  journal={Microscopy and Microanalysis},
  volume={31},
  number={Supplement\_1},
  pages={ozaf048--1066},
  year={2025},
  publisher={Oxford University Press US},
doi="10.1093/mam/ozaf048.1066"
}

@misc{crucible,
  title = {Crucible},
  authpr = {Wall, Morgan and Barnard, Edward},
  author = {{Lawrence Berkeley National Laboratory}},
  howpublished = {\url{https://crucible.lbl.gov}},
  note = {Accessed: 2025-10-09}
}

@misc{distiller2023,
  title = {Distiller},
  author = {Harris, Chris and Genova, Alessandro},
  year = {2023},
  url = {https://github.com/OpenChemistry/distiller},
  note = {Accessed: 2025-10-09}
}

@ARTICLE{gomez2018automatic,
  title     = "Automatic chemical design using a data-driven continuous
               representation of molecules",
  author    = "Gómez-Bombarelli, Rafael and Wei, Jennifer N and Duvenaud, David
               and Hernández-Lobato, José Miguel and Sánchez-Lengeling, Benjamín
               and Sheberla, Dennis and Aguilera-Iparraguirre, Jorge and Hirzel,
               Timothy D and Adams, Ryan P and Aspuru-Guzik, Alán",
  journal   = "ACS central science",
  publisher = "American Chemical Society",
  volume    =  4,
  number    =  2,
  pages     = "268--276",
  month     =  feb,
  year      =  2018,
  doi       = "10.1021/acscentsci.7b00572"
}

@ARTICLE{agrawal2016perspective,
  title     = "Perspective: Materials informatics and big data: Realization of
               the “fourth paradigm” of science in materials science",
  author    = "Agrawal, Ankit and Choudhary, Alok",
  journal   = "APL materials",
  publisher = "AIP Publishing",
  volume    =  4,
  number    =  5,
  pages     =  053208,
  month     =  may,
  year      =  2016,
  doi       = "10.1063/1.4946894"
}

@ARTICLE{midgley20033d,
  title     = "{3D} electron microscopy in the physical sciences: the
               development of {Z}-contrast and {EFTEM} tomography",
  author    = "Midgley, P A and Weyland, M",
  journal   = "Ultramicroscopy",
  publisher = "Elsevier BV",
  volume    =  96,
  number    = "3-4",
  pages     = "413--431",
  series    = "Proceedings of the International Workshop on Strategies and
               Advances in Atomic Level Spectroscopy and Analysis",
  month     =  sep,
  year      =  2003,
  doi       = "10.1016/S0304-3991(03)00105-0"
}

@article{zhou2019observing,
  title={Observing crystal nucleation in four dimensions using atomic electron tomography},
  author={Zhou, Jihan and Yang, Yongsoo and Yang, Yao and Kim, Dennis S and Yuan, Andrew and Tian, Xuezeng and Ophus, Colin and Sun, Fan and Schmid, Andreas K and Nathanson, Michael and others},
  journal={Nature},
  volume={570},
  number={7762},
  pages={500--503},
  year={2019},
  publisher={Nature Publishing Group UK London}
}

@article{turk2020promise,
  title={The promise and the challenges of cryo-electron tomography},
  author={Turk, Martin and Baumeister, Wolfgang},
  journal={FEBS letters},
  volume={594},
  number={20},
  pages={3243--3261},
  year={2020},
  publisher={Wiley Online Library}
}

@article{ribet2024uncovering,
  title={Uncovering the three-dimensional structure of upconverting core--shell nanoparticles with multislice electron ptychography},
  author={Ribet, Stephanie M and Varnavides, Georgios and Pedroso, Cassio and Cohen, Bruce E and Ercius, Peter and Scott, Mary C and Ophus, Colin},
  journal={Applied Physics Letters},
  volume={124},
  number={24},
  year={2024},
  publisher={AIP Publishing}
}

@article{chen2024imaging,
  title={Imaging interstitial atoms with multislice electron ptychography},
  author={Chen, Zhen and Shao, Yu-Tsun and Zeltmann, Steven E and Rosenberg, Ethan R and Ross, Caroline A and Jiang, Yi and Muller, David A and others},
  journal={arXiv preprint arXiv:2407.18063},
  year={2024}
}

@ARTICLE{vogl2024correlated,
  title     = "Correlated {4D}-{STEM} and {EDS} for the classification of fine
               Beta-precipitates in aluminum alloy {AA} 6063-{T6}",
  author    = "Vogl, L M and {P. Schweizer} and Donohue, J and Minor, A M",
  journal   = "Scripta materialia",
  publisher = "Elsevier BV",
  volume    =  253,
  number    =  116288,
  pages     =  116288,
  month     =  dec,
  year      =  2024,
  doi       = "10.1016/j.scriptamat.2024.116288"
}

@article{durham_scopefoundry_2018,
	title = {Scanning {Auger} spectromicroscopy using the {ScopeFoundry} software platform},
	volume = {50},
	copyright = {© 2018 John Wiley \& Sons, Ltd.},
	issn = {1096-9918},
	url = {https://onlinelibrary.wiley.com/doi/abs/10.1002/sia.6401},
	doi = {10.1002/sia.6401},
	language = {en},
	number = {11},
	urldate = {2019-02-15},
	journal = {Surface and Interface Analysis},
	author = {Durham, Daniel B. and Ogletree, D. Frank and Barnard, Edward S.},
	year = {2018},
	pages = {1174--1179},
}

@article{botifoll2024artificial,
  title={Artificial Intelligence-Assisted Workflow for Transmission Electron Microscopy: From Data Analysis Automation to Materials Knowledge Unveiling},
  author={Botifoll, Marc and Pinto-Huguet, Ivan and Rotunno, Enzo and Galvani, Thomas and Coll, Catalina and Kavkani, Payam Habibzadeh and Spadaro, Maria Chiara and Niquet, Yann-Michel and Eriksen, Martin B{\o}rstad and Mart{\'\i}-S{\'a}nchez, Sara and others},
  journal={Advanced Materials},
  pages={e06785},
  year={2024},
  publisher={Wiley Online Library},
doi="10.1002/adma.202506785"
}

\section*{Acknowledgments}

We thank Roberto dos Reis for helpful discussions.
Special thanks to Sam S Welborn and Chris Harris at the National Energy Research Scientific Computing Center for their help in connecting our MCP server to the Distiller web platform.
Work at the Molecular Foundry was supported by the Office of Science, Office of Basic Energy Sciences, of the U.S. Department of Energy under Contract No. DE-AC02-05CH11231.
This research used resources of the National Energy Research Scientific Computing Center (NERSC), a Department of Energy Office of Science User Facility using NERSC award BES-ERCAP0028035. 




\section*{Supplementary figures}

SI movie 1 focusing \label{si:movie1}

\noindent SI movie 2 tomography \label{si:movie2}

\noindent SI movie 3 Crucible + rotation series \label{si:movie3}

\noindent SI movie 4 Distiller + parallax \label{si:movie4}

\end{document}